\documentclass[aps,amsmath,amssymb,superscriptaddress,showpacs,twocolumn,floats,epsf]{revtex4}
\usepackage{graphicx,morefloats,slashed}
\usepackage{color}

\DeclareMathAlphabet{\mathpzc}{OT1}{pzc}{m}{it} 

\begin{document}
\title{Strain-induced time-reversal odd superconductivity in graphene}
\author{Bitan Roy}
\affiliation{ National High Magnetic Field Laboratory and Department of Physics, Florida State University, Florida 32306, USA}
\affiliation{ Condensed Matter Theory Center, Department of Physics, University of Maryland, College Park, Maryland 20742, USA}

\author{Vladimir  Juri\v ci\'c}
\affiliation{Institute for Theoretical Physics, Utrecht University, Leuvenlaan 4, 3584 CE Utrecht, The Netherlands}
\affiliation{Instituut-Lorentz for Theoretical Physics, Universiteit Leiden, P.O. Box 9506, 2300 RA Leiden, The Netherlands}

\begin{abstract}
{ Time-reversal symmetry breaking superconductors are exotic phases of matter with fascinating properties, which are, however, encountered rather sparsely. Here we identify the possibility of realizing such a superconducting ground state that exhibits an $f+is$ pairing symmetry in strained graphene. Although the underlying attractive interactions need to be sufficiently strong and comparable in pristine graphene to support such pairing state, we argue that strain can be conducive for its formation even for weak interactions. We show that quantum-critical behavior near the transition is controlled by a multicritical point, characterized by various critical exponents computed here in the framework of an $\epsilon$-expansion near four spacetime dimensions. Furthermore, a vortex in this mixed superconducting state hosts a pair of Majorana fermions supporting a quartet of insulating and superconducting orders, among which topologically nontrivial quantum spin Hall insulator. These findings suggest that strained graphene could provide a platform for the realization of exotic superconducting states of Dirac fermions.
}
\end{abstract}


\maketitle

The time-reversal-symmetry (TRS) breaking superconducting states are exotic phases of quantum matter, and often arise from the competition of two pairings that break distinct lattice or continuous symmetries. The proposed realizations of such states are rather sparse and some of the well-studied examples in two spatial dimensions are the $d+id$ pairing, discussed in the context of high-$T_c$ superconductivity \cite{Senthil-Marston-Fisher}, graphene\cite{Nandkishore-Levitov-Chubukov, Thomale-Abanin} and $p + ip$ pairing in fractional quantum Hall systems\cite{Read-Green}. A realization of the chiral, time-reversal odd $f$-wave pairing has been proposed in a hole-doped semiconductor, interfaced with a conventional superconductor and a magnetic insulator \cite{f+if}. Recently it has been argued that a parity and time-reversal odd \emph{axionic} $p+is$ pairing state can be realized in weakly-correlated, strong spin-orbit coupled three-dimensional doped narrow gap semiconductors \cite{Goswami-Roy}, such as $\mathrm{Sn}_{1-x}\mathrm{In}_x\mathrm{Te}$ \cite{ando-insnte} and $\mathrm{Cu}_x\mathrm{Bi}_2\mathrm{Se}_3$.

\begin{figure}[htb]
\includegraphics[width=7.00cm,height=3.50cm]{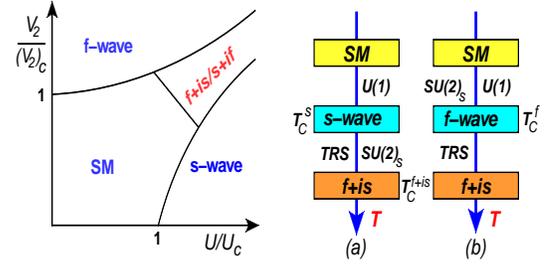}
\caption[] {Left: A schematic $T=0$ phase diagram with onsite ($U$) and next-nearest-neighbor ($V_2$) attractions. $U_c$ and $(V_2)_c$ are the critical strength of pairing interactions driving superconducting instabilities of the Dirac semimetal (SM). Right: two possible scenarios for two-stage transitions at finite temperatures: (a) when the $s$-wave transition temperature $T^s_C>T^f_C$; (b) when the $f$-wave transition temperature $T^f_C > T^s_C$ (temperature decreases in the direction of the arrow). In the cascade transitions global charge ($U(1)$), spin-rotation symmetries ($SU(2)_s$), and TRS are broken.}\label{phasediagramfis}
\end{figure}

The low-energy electron excitations in graphene, effectively described by the pseudo-relativistic Dirac equation\cite{semenoff}, can  host a plethora of broken-symmetry phases if the interactions are sufficiently strong \cite{HJR-interaction}. Interestingly enough, it is also possible, at least in principle, to realize various relativistic \emph{superconducting} orders if the net electron-electron interaction acquires an attractive component, which may be induced by the proximity effect or electron-phonon interaction\cite{igor-bitan-SC}. For example, a strong onsite (next-nearest-neighbor) attractive interaction, $U\,\, (V_2)$, supports a spin singlet(triplet) $s(f)$-wave superconductor \cite{Zhao-Paramekanti, Honerkamp}; see Fig.~1 (left). Unconventional spatially inhomogeneous superconducting states have also been proposed for the honeycomb lattice when the nearest-neighbor pairing interaction is strong enough\cite{igor-bitan-SC}.

Strain when combined with proximity effect may provide an ideal setting for realizing superconductivity in graphene. The effect of the strain or buckling can be captured by a time-reversal-symmetric axial magnetic field \cite{Igor-pseudo,GHD}, which irrespective of its spatial profile, always brings a large number of states at zero energy \cite{Igor-pseudo, Roy-Herbut-pseudo}. Henceforth, even sufficiently weak attractive interactions can give rise to pairings \cite{Ghaemi-strain, Barlas-SC}. These special, and topologically protected flat band at zero-energy in strained graphene \cite{Jackiw-pi} live on one sublattice in the bulk, while those residing on the other sublattice can only be found near the boundary of a finite graphene system \cite{Roy-Herbut-pseudo, Bitan-thesis}. Therefore, application of strain to half-filled graphene in proximity of a superconductor naturally selects only the intra-sublattice $s$- and $f$-wave pairings, since the pairing occurs exclusively among the lowest energy states which, in turn, reside on only one sublattice \cite{Zhao-Paramekanti, Honerkamp}. Remaining two fully gapped paired states, namely the Kekule superconductors \cite{igor-bitan-SC}, couple two sublattices and are therefore excluded. One can thus induce a competition between $s$- and $f$-wave pairings by placing a strained graphene flake in proximity to a regular $s$-wave, as Nb, and an unconventional odd-parity triplet superconductor, e.g., UPt$_3$, upon suppressing residual weak repulsive interactions at Dirac points\cite{HJR-interaction, igor-bitan-SC,wekcouplingexplanation}. Several questions then arise in this physical context: ($i$) What is the nature of the pairing symmetry of the ultimate superconducting state resulting from the competition of the $s$- and $f$-wave pairings? ($ii$) What is the effective field theory that captures the phase transition when these two pairings compete? ($iii$) How does the axial field influence such transition, besides being its catalyst?

We here address the competition between $s$- and $f$-wave pairings and the emergent superconducting \emph{multicriticality} using a perturbative $\epsilon (=4-d)$-expansion close to the \emph{upper critical} $d=(3+1)$ space-(imaginary-)time dimensions \cite{Zinn-Justin, Herbut-Juricic-Vafek} of the effective \emph{Gross-Neveu-Yukawa} theory. This theory contains both Lagrangians describing the quantum criticality in the vicinity of these two transitions separately, as well as incorporates the coupling of these two order-parameters(OPs). When the pairing interactions in the two channels are comparable, a TRS breaking  $f+is$ or $s+if$ state emerges at low temperatures; see Fig.~1 (right). In contrast to a pure bosonic system \cite{calebrese, Scherer, Igor-book}, transition to either of these two states is governed by a $Z_2 \otimes O(3)$ symmetric \emph{mixed} Gross-Neveu-Yukawa multicritical point due to the nontrivial Yukawa couplings, resembling in this regard the situation near an insulating quantum critical point of Dirac fermions \cite{roy-multicriticality}. The scaling of the pairing amplitudes with the axial field is essentially governed by a set of critical exponents computed in its absence. Recent surge of experimental works to realize and tune the strain-induced axial magnetic field in real and artificial graphene \cite{Levy, Manoharan, Lu-Castro-Loh}, observation of proximity induced superconductivity in graphene resting on metallic Rhenium, with $T_C \sim 2.1$K \cite{SC-graphene-rhenium} and search for the exotic broken-symmetry phases in this setup \cite{Ghaemi-strain, Roy-Herbut-pseudo, abanin-pesin} make our study important, timely, and experimentally pertinent.

The dynamics of the free Dirac fermions, living around $\pm {\bf K}$ points, where ${\bf K}=(1,1/\sqrt{3})(2\pi/a\sqrt{3})$, with $a \approx 3 \mathring{A}$ as the lattice constant, is captured by a relativistically invariant Lagrangian $L_f={\bar \Psi}(x)\sigma_0 \otimes\gamma_\mu\partial_\mu\Psi(x)$. The eight-component Dirac-Nambu spinor, invariant under the spin rotations, generated by $\vec{S}=\vec{\sigma}\otimes I_4$, is defined as $\Psi^\dagger(k)=$ $[\Psi_+^\dagger(k),\Psi_-^\dagger(k)]$, where $\Psi_\sigma^\dagger(k)= \left[ u_\sigma^\dagger(k),v_\sigma^\dagger(k),\sigma u_{-\sigma}(-k),\sigma v_{-\sigma}(-k) \right]$. Here, $k\equiv(\omega,{\bf k})$ is the three-momentum, ${\bf k}={\bf K}+{\bf q}$, $|{\bf q}| \ll |{\bf K}|$, and summation over repeated indices is assumed. $\sigma=\pm$ are the spin projections along the $z$-axis, $u_\sigma,v_\sigma$ are the spinor components on the two sublattices, and as usual ${\bar\Psi}\equiv\Psi^\dagger\sigma_0 \otimes \gamma_0$ \cite{Herbut-topology}. The real ($\varphi$) and the imaginary ($\chi$) parts of the $s$-wave  order parameter (OP) are defined as
\begin{eqnarray}
\Phi(x)=\langle{\bar\Psi}(x)\sigma_0 \otimes ( I_4 c \theta + i\gamma_5 s \theta)\Psi(x)\rangle \equiv \varphi(x)+i\chi(x),
\end{eqnarray}
after rotating the spinor as $\Psi\rightarrow U\Psi$ with $U=\exp[i\frac{\pi}{4}\sigma_0\otimes\gamma_3]$, $c\equiv \cos$ and $s\equiv \sin$. Such rotation leaves $L_f$ invariant and allows the extension of the theory from the physical $(2+1)$- to $(3+1)$-dimensions. The coupling of the $s$-wave OP with the gapless Dirac fermions assumes the form of the Yukawa interaction
\begin{equation}
L_{b-f}^s=g_{1s} \; \varphi\left( {\bar\Psi}\sigma_0\otimes I_4\Psi \right) + g_{2s}\; \chi\left({\bar\Psi}\sigma_0\otimes i\gamma_5\Psi\right),
\end{equation}
with $g_{1s/2s}\sim U$. The effective theory describing the transition into the $s$-wave superconductor is given by $L^s=L_f+L_{b-f}^s+L^s_b$, with
\begin{equation}\label{singlet-GL}
L^s_b= \sum_{\alpha=\varphi,\chi}\bigg[ \frac{1}{2}\left( \partial_\mu \alpha \right)^2 + m^2_\alpha \alpha^2 +\frac{\lambda_\alpha}{4!} \alpha^4 \bigg]
+\frac{\lambda_{\varphi \chi}}{12} \varphi^2 \chi^2,
\end{equation}
as the Ginzburg-Landau Lagrangian describing the dynamics of the singlet OP. The superconducting mass $m^2_\alpha \sim (U-U_c)$, where $U_c$ is the zero-axial-field critical onsite attraction for $s$-wave ordering. We here allow a generic situation where the Yukawa and the bosonic couplings for the real and the imaginary parts of the OP are different. As we show near the multicritical point for the transition into the $f+is$ ($s+if$) state, only one Yukawa and bosonic quartic couplings are nontrivial.

Analogously, the effective theory $L^t=L_f+L^t_{b-f}+L^t_b$, defined below, describes the universal behavior near the transition into the triplet $f$-wave paiting. The real $(\vec{\varphi})$ and imaginary $(\vec{\chi})$ part of the $f$-wave OP $\vec{\Phi}(x) \equiv \vec{\varphi}(x)+ i \vec{\chi}(x)$, is obtained from Eq.~(1) by replacing $\sigma_0$ by $\vec{\sigma}$. This OP is even (odd) under the sublattice (valley) exchange \cite{igor-bitan-SC, Herbut-topology}, and its coupling with the Dirac fermions again assumes the Yukawa form
\begin{equation}
L_{b-f}^t=g_{1t}\; {\vec\varphi} \cdot \left( {\bar\Psi}{\vec\sigma}\otimes I_4 \Psi \right) + g_{2t}\; {\vec\chi} \cdot \left({\bar\Psi}{\vec\sigma}\otimes i\gamma_5\Psi\right),
\end{equation}
where $g_{1t/2t}\sim V_2$. The dynamics of the triplet OP is described by the Lagrangian
\begin{equation}
L_b^t= \sum_{{\vec\alpha}={\vec\varphi},{\vec\chi}}\bigg[ \frac{1}{2}(\partial_\mu{\vec\alpha})^2+m^2_{\vec{\alpha}}{\vec\alpha}^2 + \frac{\lambda_{{\vec\alpha}}}{4!} \left(\vec{\alpha} \cdot \vec{\alpha} \right)^2 \bigg] + \frac{\lambda_{{\vec\varphi}{\vec\chi}}}{12}{\vec\varphi}^2{\vec\chi}^2,
\end{equation}
a generalization of the singlet version in Eq.~(\ref{singlet-GL}), with $m^2_{\vec{\alpha}} \sim V_2-(V_2)_c$, where $(V_2)_c$ is the zero-axial-field critical interaction for $f$-wave pairing. Also here we allow for all the bare couplings to be different.

The coupling of the singlet and the triplet OPs close to the multicritical point is given by
\begin{equation}
L_b^{st}=\frac{\lambda_{\varphi{\vec\varphi}}}{12}\varphi^2{\vec\varphi}^2+\frac{\lambda_{\chi{\vec\chi}}}{12}\chi^2{\vec\chi}^2
+\frac{\lambda_{\varphi{\vec\chi}}}{12}\varphi^2{\vec\chi}^2+\frac{\lambda_{\chi{\vec\varphi}}}{12}\chi^2{\vec\varphi}^2,
\end{equation}
and their competition is described by the Lagrangian $L=L_f+L_b^s+L_{b-f}^s+L_b^t+L_{b-f}^t+L_{b}^{st}$. Throughout the Letter, we omit the couplings of the massless Dirac fermions and the superconducting OPs with the \emph{fluctuating gauge fields}, since both the Fermi and the bosonic velocities are much smaller that the velocity of light. The ultimate critical behavior is, however, governed by a \emph{charged critical point}, where all the velocities are equal, although such deep infrared critical behavior may not be accessible experimentally due to their logarithmically slow increase \cite{RJH-RC}.

Since all the Yukawa and the quartic bosonic couplings in this theory are {\it exactly} marginal in $d=(3+1)$, we use the $\epsilon$-expansion about four dimensions, with $\epsilon=4-d$, as the tool for studying the quantum-critical behavior. The standard minimal-subtraction scheme \cite{Zinn-Justin, supplementary}, yields the (infrared) $\beta$-functions for the Yukawa couplings $g_{1s}$, $g_{1t}$
\begin{eqnarray}
\beta_{g_{1s}^2} &=& \epsilon g_{1s}^2-(2N+3)g_{1s}^4+g_{1s}^2(g_{2s}^2-9g_{1t}^2+3g_{2t}^2),\nonumber\\
\beta_{g_{1t}^2} &=& \epsilon g_{1t}^2-(2N+1)g_{1t}^4-g_{1t}^2(5g_{2t}^2+3g_{1s}^2-g_{2s}^2),
\end{eqnarray}
after taking $N_d g^2\rightarrow g^2$, where $N_d\equiv S_d/(2\pi)^d$ and $S_d\equiv 2\pi^{d/2}/\Gamma(d/2)$, with $N (=2$ for graphene) as the number of four-component Dirac fermions. The $\beta$-functions for the other two Yukawa couplings, $g_{2s}$ and $g_{2t}$, are obtained by replacing $1\leftrightarrow2$ in the above two $\beta$-functions. Interestingly, the $\beta$-functions for the Yukawa couplings are decoupled from the bosonic quartic couplings ($\lambda_\alpha$), and out of 16 fixed points in the four-dimensional subspace spanned by the Yukawa couplings, only \emph{two} are \emph{fully stable}. One that describes the transition into the $f+is$ state is located at
\begin{equation}
\label{Yukawa-f+is}
(g^\ast_{2s})^2=(g^{\ast}_{1t})^2=\frac{\epsilon}{2 N}, \: g^\ast_{1s}=g^\ast_{2t}=0,
\end{equation}
while the other one that corresponds to the transition into the $s+if$ state is obtained by replacing $1\leftrightarrow2$ above. Therefore, our renormalization group analysis suggests that when singlet and triplet pairings compete in the system of massless Dirac fermions, the transition is {\it always} multicritical in nature and towards the formation of a mixed $f+is$ or $s+if$ state, with TRS dynamically broken due to the Yukawa interactions; see Fig.~1.

This outcome can further be substantiated from the minimization of free-energy. The effective single-particle Hamiltonian with both singlet and triplet OPs reads
\begin{equation}\label{single-particle}
H=\sigma_0\otimes i\gamma_0{\vec \gamma}\cdot {\vec p}+\sum_{x=s,f} \Delta_x \sigma_x \otimes\gamma_0(\cos\theta_x +i\gamma_5 \sin\theta_x),
\end{equation}
where $\sigma_{s/f}=\sigma_{0/3}$, ${\vec p}$ is the momentum operator and we fix the spin quantization of the $f$-wave OP along the $z$-axis, for simplicity. Its spectrum contains two branches of positive energy
\begin{equation}
E_\pm=\sqrt{p^2+\Delta_s^2+\Delta_f^2\pm2\Delta_s\Delta_f\cos(\theta_s-\theta_f)},
\end{equation}
and the corresponding negative ones at $-E_\pm$. Therefore, the energy of the filled Dirac-Fermi sea at half filling $-(E_+ + E_-)$, is maximally lowered when $\theta_s-\theta_f=\pm \pi/2$. All the terms in Eq.~(\ref{single-particle}) then enter as the sum of the squares in the expression of energy, and the Dirac points are maximally gapped. This constraint corresponds to two exactly degenerate TRS breaking $f+is$ and $s+if$ states, as we have found from the above renomalization group calculation, suggesting its robustness against fluctuations. In turn, this gives confidence that our result could be valid beyond the leading order in the $\epsilon-$expansion.

Near the Yukawa fixed point in Eq.~(\ref{Yukawa-f+is}), where only the imaginary (real) part of the $s(f)$-wave order-parameter is nonvanishing, three bosonic quartic couplings $(\lambda_\chi,\lambda_{{\vec \varphi}},\lambda_{\chi{\vec\varphi}})$ are nontrivial, yielding a $Z_2\otimes O(3)$ symmetric critical theory. Their $\beta$-functions yield only one fully stable fixed point, which for $N=2$ (graphene) in the critical plane ($m_\chi=m_{\vec\varphi}=0$) is located at $(\lambda^\ast_{\chi},\lambda^\ast_{{\vec \varphi}},\lambda^\ast_{\chi{\vec\varphi}})=(0.972,0.992,1.103)\epsilon$ \cite{supplementary}. Hence, the transition to the $f+is$ ($s+if$) state is described by a \emph{fermionic mixed critical point}, where besides the Yukawa couplings of the singlet and the triplet OPs, their quartic interactions as well as their mutual coupling are finite. In the absence of Yukawa couplings, two fully stable fixed points in a $Z_2\otimes O(3)$ purely bosonic theory are \emph{decoupled} ($\lambda_{\chi{\vec\varphi}}=0$). They are located at $\left( \lambda_\chi,\lambda_{{\vec\varphi}} \right)=(2\epsilon/3,0)$ [$(0,6\epsilon/11)$], and describe the transition into the pure singlet [triplet] phase. Since the anomalous dimension of the bosonic fields $\eta_{\vec{\varphi}}(=0.020\epsilon^2) > \eta_{\chi}(=0.018\epsilon^2)$ \cite{Igor-book}, the ultimate criticality in the purely bosonic theory is possibly governed by the triplet critical point \cite{Zinn-Justing-etaconjecture}. In contrast, the massless fermions, through Yukawa couplings to the critical bosonic fluctuations, stabilize the mixed multicritical point in graphene governing the transition into the $f+is$ ($s+if$) state. The bicritical fixed points in our theory lie in the unphysical regime of couplings $(\lambda 's<0)$. Therefore, the transition into $f+is$ ($s+if$) state is always continuous in nature.

The critical theory possesses two relevant operators, the masses $m_\chi$ and $m_{\vec\varphi}$, that tune the phase transition into the mixed state. Their flow defines the correlation-length exponents $(\nu_\chi,\nu_{\vec\varphi})=(1/2+0.509\epsilon,1/2+0.521\epsilon)$ in the vicinity of the above multicritical point. The anomalous dimensions for spin-singlet/triplet OP and Dirac fermions close to the transition is $\eta_{\chi/\vec{\varphi}}=\epsilon$ and $\eta_\Psi=\epsilon/N$, respectively \cite{supplementary}. Since the mass in the $f+is$ state is Lorentz-symmetric, we expect weak Lorentz-symmetry-breaking perturbations to be irrelevant and the dynamical critical exponent $z=1$, close to this critical point \cite{Herbut-Juricic-Vafek}. The residue of the quasiparticle pole vanishes as $m_\alpha^{z\nu_\alpha\eta_\Psi}$ with $\alpha=\chi,{\vec\varphi}$, depending on the relevant direction from which  $f+is$ state is approached, and Dirac fermions cease to exist as sharp quasiparticle excitations at the transition. The critical exponents near the superconducting multicritical point are different than the ones in the vicinity of the pure $s$-wave transition, where the correlation-length exponent is $\nu=1/2+0.3 \epsilon$, and fermionic (bosonic)  anomalous dimension is $\eta_{\Psi}=4\epsilon/6$ ($\eta_b=\epsilon/6$)\cite{RJH-RC}.

The quantum-critical behavior near the transition into a pure $f$-wave superconducting state in graphene is described by a critical point located at $(g_{t}^2,\lambda_t)=(0.1,0.498) \epsilon$, where $g_t \equiv g_{1t}=g_{2t}$ and $\lambda_t \equiv \lambda_{\vec{\varphi}}=\lambda_{\vec{\chi}}=\lambda_{\vec{\varphi} \vec{\chi}}$. The critical exponents near this critical point are found to be  $\nu=1/2+0.266 \epsilon$, $\eta_\Psi=3\epsilon/10$, and $\eta_b=2\epsilon/5$ and are of a distinct non-mean-field nature \cite{supplementary}.

As the temperature is gradually lowered, the system first enters the dominant paired $s$($f$)-wave state at temperature $T_C^s$($T_C^f$); see Fig.\ 1(right). Only at even lower temperature ($T_C^{f+is}$) the $f+is$ state is reached with TRS being broken, which can be confirmed by Kerr rotation measurements \cite{Kapitulnik}. This two-stage superconducting transition leads to discontinuities in the specific heat at the two critical temperatures. Since the $f$-wave pairing is also more susceptible to generic disorders, it is also possible that the transition temperature for the $s$-wave pairing is higher than that for the $f$-wave pairing\cite{disorder-1}, which is devoid of Pauli limiting field\cite{tinkham}. Consequently, the lower critical field $H_{c1}$ discerns abrupt increment below the second transition temperature $T_C^{f+is}$. Since spin of the $f$-wave order can be flipped by applying a radio frequency signal in a nuclear magnetic resonance experiment, after switching off this signal, as the spin relaxes back to the ground state, a radio frequency signal is emitted.

The above critical exponents govern the scaling of the physical quantities, such as the pairing gap, in the presence of axial fields, which catalyze the pairings without changing the universality class of the pertaining transitions. In weak axial fields, the pairing gap ($\Delta$) exhibits the scaling
\begin{equation}
\frac{\Delta}{g} =(l_b/a)^{-\beta/\nu} F \left[(l_b/a)^{1/\nu} \; (g-g_c) \right],
\end{equation}
similar as in a finite system of the size $l_b$, where the axial magnetic length $l_b=\sqrt{\hbar/eb} \sim 100 a$ for $b \sim 100$T, and therefore the continuum description remains justified. Here, $\beta$ is the OP exponent,  $g-g_c$ measures the deviation from the zero-field critical interaction ($g_c$), and $F(x)$ is the scaling function \cite{Igor-Bitan-scaling, Fakher-Igor}. Standard relations between the critical exponents then yield $2\beta/\nu=d+z-2+\eta_b$ $=2-(\epsilon-\eta_b)$, in effective $d+z=4-\epsilon$ dimensions, with $\eta_b$ as the anomalous dimension of the OP at the zero-field critical point \cite{Igor-book}. Different critical exponents near the transition into $s$-wave, $f$-wave, and the $f+is$ states, therefore, lead to distinct scalings of the gap and concomitantly their transition temperatures with axial fields, which should serve as a clear signal of the cascade two-stage transition, Fig.\ 1 (right). Even for weak enough pairing interactions, $\Delta \sim g b \sim T_C$(BCS approximation) in the absence of disorder \cite{Igor-Bitan-scaling}, resulting in sizable pairing gaps when $b \sim 50-300$T \cite{Levy, Lu-Castro-Loh, Manoharan}. Hence, the proposed unconventional $f+is$ state can possibly be realized in cleaner graphene samples, with $T^{f+is}_C \sim 1$K, as a recent experiment would suggest \cite{SC-graphene-rhenium}.

Type-II nature of the superconductors inducing the mixed $f+is$ state in graphene through proximity allows us to investigate the vortex phase therein. Since the single-particle Hamiltonian in Eq.\ (\ref{single-particle}) in the $f+is$ state represents \emph{two} copies of \emph{anisotropic} Jackiw-Rossi Hamiltonian \cite{Jackiw-Rossi}, a single vortex hosts two Majorana fermions \cite{Herbut-topology}, and supports \emph{four} Dirac masses, $\left\{\sigma_3 \otimes \gamma_0, \sigma_0 \otimes \gamma_0, \sigma_1 \otimes i \gamma_0 \gamma_3, \sigma_2 \otimes i \gamma_0 \gamma_3 \right\}$, inside the vortex core\cite{Herbut-topology}, even when real and the axial magnetic fields are present simultaneously \cite{Roy-realpseudo-zeromode}. All these four masses anticommute with $H$ in Eq.\ (\ref{single-particle}), and leave the zero energy subspace invariant. They respectively represent the charge-density wave (CDW), $z$-component of the topological spin Hall insulator, and the remaining spin components of the $f$-wave superconductor. It is therefore conceivable to realize $f$-wave pairing inside the vortex core, however with different spin components than in the bulk. On the other hand, the Zeeman coupling, $H_Z=g B(\vec{x}) \left( \sigma_3 \otimes I_4 \right)$, where $g \approx 2$ in graphene and $B(\vec{x})$ representing the magnetic field, supports CDW inside the vortex core of the $f+is$ state, which can be detected using scanning tunneling microscopy, for instance. In the $f$-wave phase the Majorana states support the easy axis components of the N\'{e}el OP, represented by the matrices $\left( \sigma_1, \sigma_2 \right) \otimes i \gamma_1 \gamma_2$, besides the CDW and the quantum spin Hall insulator; latter is then favored by the Zeeman coupling \cite{Herbut-topology}.

To summarize, we here demonstrated the possibility of realizing a novel TRS breaking paired ground state with $f+is$ symmetry in strained graphene, and addressed its universal properties. Moreover, our results may be consequential for other condensed-matter systems exhibiting low-energy Dirac quasiparticles, such as structurally similar monolayer silicene \cite{silicene-SC}, surface states of topological insulators, and Weyl semimetals.

The authors would like to acknowledge fruitful discussions with Oskar Vafek, Igor Herbut, and Pallab Goswami. B. R. was supported at the National High Magnetic Field Laboratory by NSF Cooperative Agreement No.DMR-0654118, the State of Florida, and the U. S. Department of Energy. V. J. acknowledges the support of the Netherlands Organization for Scientific Research (NWO). The authors are grateful to Advance Study Group and workshop ``Spin Orbit Entanglement: Exotic States of Quantum Matter in Electronic Systems" at the Max-Planck Institute for Complex Systems, Dresden, Germany, where part of this work was finalized.

\newpage

\onecolumngrid

\vspace{1.5cm}

\begin{center}
{\bf \large Supplementary material for ``Strain-induced time-reversal odd superconductivity in graphene''} \\
{Bitan Roy$^{1,2}$ and Vladimir Juri\v ci\' c$^{3,4}$ } \\
{$^1$ National High Magnetic Field Laboratory, Florida State University, Tallahassee, Florida 32306, USA}\\
{$^2$Condensed Matter Theory Center, Department of Physics, University of Maryland, College Park, Maryland 20742, USA} \\
{$^3$Institute for Theoretical Physics, Utrecht University, Leuvenlaan 4, 3584 CE Utrecht, The Netherlands}\\
{$^4$ Instituut-Lorentz for Theoretical Physics, Universiteit Leiden, P.O. Box 9506, 2300 RA Leiden, The Netherlands }\\
\end{center}

We here present details of the renomalization group calculation to the one-loop order using the Euclidean partition function. \cite{justin}
\section{ Computation of bosonic self-energy}
\begin{figure}[htb]\label{boson-self-energy}
\includegraphics[width=9.0cm,height=2.5cm]{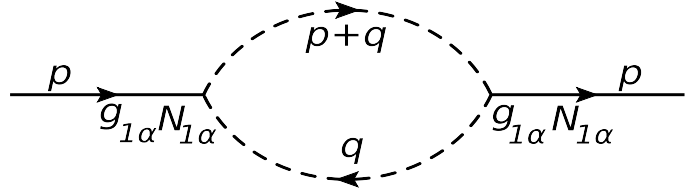}
\caption{One loop correction to the bosonic self-energy}
\end{figure}

Each of the bosonic order-parameters (OPs), ${\varphi},\chi,{\vec\varphi},{\vec\chi}$, gets renormalized due to the Yukawa coupling $g_{1s},g_{2s},g_{1t},g_{2t}$, respectively, and in full generality we can only consider the real part of the OPs, $\varphi$ and ${\vec\varphi}$. Then the diagram in Fig. 2 yields
\begin{equation}
(1)=g_{1\alpha}^2S_{\alpha} Tr\int\frac{d^d q}{(2\pi)^d}\frac{N_{1\alpha}\slashed{q}N_{1\alpha}(\slashed{q}+\slashed{p})}{q^2(q+p)^2}=g_{1\alpha}^2S_{\alpha} Tr\int\frac{d^d q}{(2\pi)^d}\frac{\slashed{q}(\slashed{q}+\slashed{p})}{q^2(q+p)^2}=-g_{1\alpha}^2\,\,S_\alpha\,\, 2N\,\,p^2\,\, \frac{N_d}{\epsilon},
\end{equation}
with $\alpha=s,t$, $(N_{1s},N_{1t})=(\sigma_0\otimes I_4,{\vec\sigma\otimes I_4} )$, $S_{s,t}=1,3$ for the singlet and the triplet OPs, respectively, $\slashed{q}\equiv \gamma_\mu q_\mu$, and $N_d\equiv S_d/(2\pi)^d$ with $S_d\equiv 2\pi^{d/2}/\Gamma(d/2)$. Furthermore, bosonic quartic couplings contribute only to the mass renormalization to the one-loop order. For instance, the mass of the real singlet OP is renormalized by the vertices $\lambda_\varphi,\lambda_{\varphi\chi},\lambda_{\varphi{\vec\varphi}}$, and $\lambda_{\varphi{\vec\chi}}$. Taking into account combinatorial factors for these diagrams and the above fermion contribution, we arrive at the renormalization conditions for the propagators of the real singlet and the triplet OPs
\begin{eqnarray}
&&Z_\varphi(p^2+m_{\varphi,0}^2)+2N g_{1s}^2\frac{1}{\epsilon}p^2-\left(\frac{\lambda_\varphi}{2}+\frac{\lambda_{\varphi\chi}}{6}+\frac{\lambda_{\varphi{\vec\varphi}}}{2}+
\frac{\lambda_{\varphi{\vec\chi}}}{2}\right)\frac{1}{\epsilon}m_{\varphi}^2=p^2+m_\varphi^2,\\
&&Z_{\vec\varphi}(p^2+m_{{\vec\varphi},0}^2)+2N g_{1t}^2\frac{1}{\epsilon}p^2-\left(\frac{5\lambda_{\vec\varphi}}{6}+\frac{\lambda_{{\vec\varphi}{\vec\chi}}}{2}
+\frac{\lambda_{\varphi{\vec\varphi}}}{6}+
\frac{\lambda_{\chi{\vec\varphi}}}{6}\right)\frac{1}{\epsilon}m_{{\vec\varphi}}^2=p^2+m_{\vec\varphi}^2,
\end{eqnarray}
while the analogous renormalization conditions for the fields $\chi,{\vec\chi}$ are obtained from the above equations by replacing $\{\varphi,{\vec\varphi}\}\leftrightarrow \{ \chi,{\vec\chi} \}$, and $1\rightarrow2$ in the Yukawa couplings.  Here, we redefined the couplings as $N_d Q\rightarrow Q$ where $Q=\{g^2,\lambda\}$, and the parameters with the subscript ``0" are the bare ones. These conditions then yield
\begin{equation}\label{Z-bosonic}
Z_\varphi=1-2Ng_{1s}^2\frac{1}{\epsilon},\,\,
Z_{\vec\varphi}=1-2Ng_{1t}^2\frac{1}{\epsilon},
\end{equation}
and the analogous expressions for the fields $\chi,{\vec\chi}$ are obtained by replacing $1\rightarrow2$. Renormalization conditions for the masses are also readily obtained
\begin{eqnarray}
&&Z_\varphi m_{\varphi,0}^2\mu^{-\epsilon}-\left(\frac{\lambda_\varphi}{2}+\frac{\lambda_{\varphi\chi}}{6}+\frac{\lambda_{\varphi{\vec\varphi}}}{2}+
\frac{\lambda_{\varphi{\vec\chi}}}{2}\right)\frac{1}{\epsilon}m_{\varphi}^2=m_{\varphi}^2,\\
&&Z_{\vec\varphi}m_{{\vec\varphi},0}^2\mu^{-\epsilon}-\left(\frac{5\lambda_{\vec\varphi}}{6}+\frac{\lambda_{{\vec\varphi}{\vec\chi}}}{2}
+\frac{\lambda_{\varphi{\vec\varphi}}}{6}+
\frac{\lambda_{\chi{\vec\varphi}}}{6}\right)\frac{1}{\epsilon}m_{{\vec\varphi}}^2=m_{\vec\varphi}^2,
\end{eqnarray}
with $\mu$ being the renormalization scale.

\section{Self-energy of the Dirac fermions}
\begin{figure}[htb]\label{fermion-self-energy}
\includegraphics[width=9.0cm,height=2.5cm]{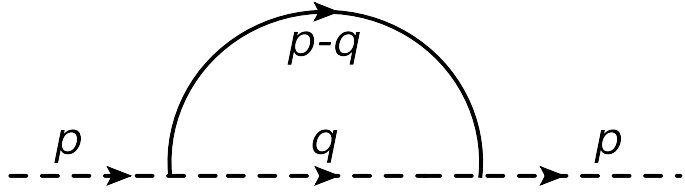}
\caption{One loop correction to the self-energy of the fermions}
\end{figure}

Dirac fermions receive self-energy corrections from all the four OPs through the Yukawa couplings. Singlet OP field yields a correction (Fig. 3)
\begin{equation}
(2)=(g_{1s}^2+g_{2s}^2)N_s\int\frac{d^d q}{(2\pi)^d}\frac{i\slashed{q}}{q^2}\frac{1}{(q-p)^2}=(g_{1s}^2+g_{2s}^2)N_s\left(\frac{1}{2\epsilon}\right)i\slashed{p}N_d,
\end{equation}
with $N_s=1$, while  the contribution arising from the Yukawa coupling in the triplet channel is obtained by replacing $s\rightarrow t$ and using $N_t=3$ as the number of the components of the triplet OPs. The total contribution then yields the wavefunction renormalization of the Dirac field to the one-loop order
\begin{equation}\label{Z-Psi}
Z_\Psi=1-\frac{1}{2}(g_{1s}^2+g_{2s}^2+3g_{1t}^2+3g_{2t}^2)\frac{1}{\epsilon}.
\end{equation}

\section{Renormalization of the Yukawa vertices}
\begin{figure}[htb]\label{yukawa-vertex}
\includegraphics[width=6.0cm,height=4.5cm]{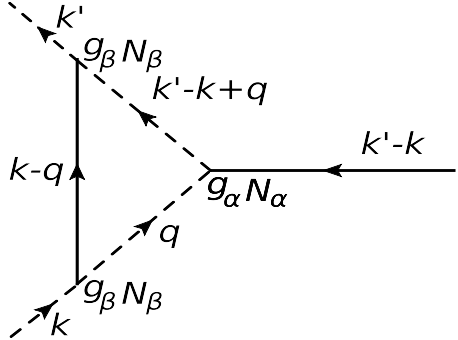}
\caption{One loop correction to the Yukawa vertex.}
\end{figure}

Four Yukawa vertices are given by the matrices $N=\{\sigma_0\otimes I_4,\sigma_0\otimes i\gamma_5,{\vec\sigma}\otimes I_4,{\vec\sigma}\otimes i\gamma_5\}$ and correspond to the couplings $g_\alpha$ with $\alpha=\{{1s},{2s},{1t},{2t}\}$, respectively. The renormalization of the Yukawa vertex with the matrix $N_\alpha$ and the coupling $g_\alpha$ arising from the Yukawa coupling with the vertex containing the matrix $N_\beta$ and the coupling constant $g_\beta$ is given by the diagram in Fig. 4 and has the form
\begin{equation}
g_{\alpha\beta}=g_\alpha g_\beta^2\int\frac{d^d q}{(2\pi)^d}N_\beta\frac{i\slashed{q}}{q^2}N_\alpha\frac{i(\slashed{q}+\slashed{k}'-\slashed{k})}{(q+k'-k)^2}N_\beta\frac{1}{(q-k)^2},
\end{equation}
with no summation over the repeated indices $\alpha,\beta$ assumed.
Setting $k'=0$ and using the identities $\{\gamma_5,\gamma_\mu\}=0$, ${\vec\sigma}\cdot{\vec\sigma}=3$ and $\sigma_i\sigma_j\sigma_i=-\sigma_j$, we obtain
 \begin{equation}
 g_{\alpha\beta}=-C_{\alpha\beta}g_\alpha N_\alpha g_\beta^2 I,
 \end{equation}
  with
 \begin{equation}
 I=\int\frac{d^d q}{(2\pi)^d}\frac{i\slashed{q}}{q^2}\frac{i(\slashed{q}-\slashed{k})}{(q-k)^4}=-\frac{N_d}{\epsilon}+O(1),
 \end{equation}
and the coefficients $C_{\alpha\beta}$ given in Table I. This finally yields
\begin{equation}\label{C-matrix}
g_{\alpha\beta}=C_{\alpha\beta}g_\alpha g_\beta^2 N_d\frac{1}{\epsilon}.
\end{equation}
Renormalization conditions for the couplings $g_{1s}$ and $g_{1t}$ then read
\begin{eqnarray}
&&Z_\varphi^{1/2} Z_\Psi g_{1s,0}\mu^{-\epsilon}+g_{1s}(-g_{1s}^2+g_{2s}^2-3g_{1t}^2+3g_{2t}^2)\frac{1}{\epsilon}=g_{1s},\\
&&Z_{\vec\varphi}^{1/2} Z_\Psi g_{1t,0}\mu^{-\epsilon}+g_{1t}(-g_{1s}^2+g_{2s}^2+g_{1t}^2-g_{2t}^2)\frac{1}{\epsilon}=g_{1t}
\end{eqnarray}
while the renormalization conditions for the remaining two couplings $g_{2s}$ and $g_{2t}$ are obtained from above equations by replacing $1\leftrightarrow2$, and all the couplings are redefined as $N_d g^2\rightarrow g^2 $ here. These renormalization conditions, using Eqs.\ (\ref{Z-bosonic}) and (\ref{Z-Psi}), yield the beta functions for the Yukawa couplings shown in the main text. For a coupling $Q$, the infrared beta-function is $\beta_Q=-\beta_Q^{\rm UV}=-dQ/d\ln\mu$, where $\beta_Q^{\rm UV}$ is the ultraviolet beta-function, and $\mu$ is the renormalization energy scale\cite{justin}.
\begin{table}[h]
  \begin{tabular}{|c||c|c|c|c|}
    \hline
   $\alpha$\textbackslash$\beta$  & {$1s$} & {$2s$} & {$1t$} & {$2t$} \\
   \hline \hline
    {$1s$} & {-1} & 1 & -3 & 3 \\
   \hline
    {$2s$} & 1 & -1 & 3 & -3  \\
   \hline
    {$1t$} & -1 & 1 & 1 & -1 \\
   \hline
    {$2t$} & 1 & -1 & -1 & 1 \\
    \hline
  \end{tabular}
  \caption{Coefficients $C_{\alpha\beta}$ in Eq.\ (\ref{C-matrix}) giving the correction to the Yukawa vertex $g_\alpha$ arising from the Yukawa coupling $g_\beta$. }
  \end{table}

\section{Renormalization of the bosonic quartic couplings}
\begin{figure}[htb]\label{bosonic-vertex}
\includegraphics[width=6.0cm,height=4.5cm]{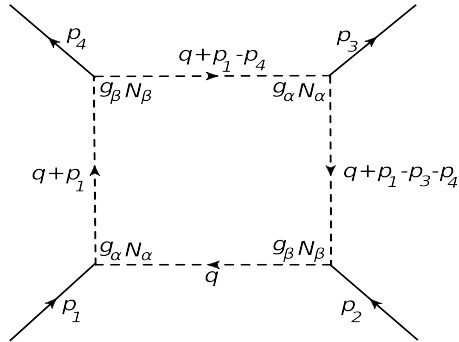}
\caption{Correction to the quartic bosonic couplings due to Yukawa interactions.}
\end{figure}

Each of the quartic couplings in the theory receive a correction from the fermions due to the Yukawa interactions, see Fig. 5. Since each quartic coupling contains at most two different pairs of the bosonic fields, in general, the vertex correction for the coupling $\lambda_{\alpha\beta}$ with $\alpha,\beta=\{\varphi,\chi,{\vec\varphi},{\vec\chi}\}$ ($\lambda_{\alpha\alpha}\equiv\lambda_\alpha$ and $\lambda_{\alpha\beta}=\lambda_{\beta\alpha}$) has the form
\begin{equation}\label{quartic-Yukawa correction}
{\tilde\delta}\lambda_{\alpha\beta}=6 \; g_\alpha^2 g_\beta^2  \;Tr \; \int \; \frac{d^d q}{(2 \pi)^d} \: \frac{N_\alpha \; \slashed{q}\; N_\beta \left( \slashed{p}_1 +\slashed{q}-\slashed{p}_3-\slashed{p}_4 \right) \; N_\alpha\;
\left( \slashed{p}_1 + \slashed{q}-\slashed{p}_4 \right) \; N_\beta \; \left( \slashed{p}_1 + \slashed{q}\right)}
{q^2 \; (p_1+q)^2 \; (p_1+q-p_4)^2 \; (p_1+q-p_3-p_4)^2} = 24 g_\alpha^2 g_\beta^2 N\frac{N_d}{\epsilon},
\end{equation}
with $g_\varphi\equiv g_{1s},g_\chi\equiv g_{2s}, g_{\vec\varphi}\equiv g_{1t}$, and $g_{\vec\chi}\equiv g_{2t}$.
\subsection{Singlet channel}
The quartic couplings are renormalized also from the bosonic fields. The vertex corrections are  given by the standard expressions\cite{justin}. For instance, quartic coupling for the real part of the singlet OP, $\lambda_\varphi$, receives corrections from  all the vertices that contain $\varphi$ in the index, i.e.,  $\lambda_{\varphi},\lambda_{\varphi\chi},\lambda_{\varphi{\vec\varphi}}$, and $\lambda_{\varphi{\vec\chi}}$. Straightforward calculation of the combinatorial factors (all the loop integrals are equal to $N_d/\epsilon+O(1)$) gives
\begin{equation}
\delta\lambda_\varphi=-\frac{N_d}{\epsilon}\left(\frac{3}{2}\lambda_\varphi^2+\frac{1}{6}\lambda_{\varphi\chi}^2
+\frac{1}{2}\lambda_{\varphi{\vec\varphi}}^2+\frac{1}{2}\lambda_{\varphi{\vec\chi}}^2\right),
\end{equation}
and the vertex correction for $\lambda_\chi$ is obtained by replacing $\varphi\leftrightarrow\chi$ above.
Vertex correction for the mixed coupling $\lambda_{\varphi\chi}$ is given by
\begin{equation}
\delta\lambda_{\varphi\chi}=-\frac{N_d}{2\epsilon}\left(\lambda_{\varphi\chi}(\lambda_\varphi+\lambda_\chi)+\frac{4}{3}\lambda_{\varphi\chi}^2
+\lambda_{\varphi{\vec\varphi}}\lambda_{\chi{\vec\varphi}}+\lambda_{\chi{\vec\chi}}\lambda_{\varphi{\vec\chi}}\right).
\end{equation}
Notice that when $\lambda_{\varphi}=\lambda_\chi=\lambda_{\varphi\chi}\equiv \lambda_s$ and all the other quartic couplings set to zero, the quartic coupling for the singlet OP, given in Eq.\ (3) in the main text, acquires the form $\lambda(\varphi^2+\chi^2)^2$, and from the above equations the vertex corrections for all the three bosonic couplings in the singlet channel become equal, $\delta\lambda_s=\frac{5}{3}\lambda_s^2$, as they should be with the coefficient that agrees with the known value. \cite{justin, RJH-RC1} Moreover, when also taking into account the limit when the two Yukawa couplings in the singlet channel are equal, $g_{1s}=g_{2s}\equiv g_s$ and the ones in the triplet channel set to zero, one obtains already known beta functions for the effective Gross-Neveu-Yukawa theory for the pure $s$-wave superconducting transition derived in Ref.\ \onlinecite{RJH-RC1} after the couplings are redefined as $\lambda\rightarrow 3\lambda$ and $g\rightarrow g/\sqrt{2}$.
\subsection{Triplet channel}

Renormalization of the $\lambda_{{\vec\varphi}}$-vertex arises from all the couplings containing ${\vec\varphi}$ in the index and straightforward computation yields
\begin{equation}
\delta\lambda_{\vec\varphi}=-\frac{N_d}{\epsilon}\left(\frac{11}{6}\lambda_{\vec\varphi}^2+\frac{1}{2}\lambda_{{\vec\varphi}{\vec\chi}}^2
+\frac{1}{6}(\lambda_{\varphi{\vec\varphi}}^2+\lambda_{\chi{\vec\varphi}}^2)\right),
\end{equation}
while the analogous correction for the $\lambda_{\vec\chi}$-vertex is obtained by replacing ${\vec\varphi}\leftrightarrow{\vec\chi}$ in the above equation.
Correction to the $\lambda_{{\vec\varphi}{\vec\chi}}$-vertex coming from the quartic couplings reads
\begin{equation}
\delta\lambda_{{\vec\varphi}{\vec\chi}}=-\frac{N_d}{\epsilon}\left(\frac{5}{6}(\lambda_{\vec\varphi}+\lambda_{\vec\chi})\lambda_{{\vec\varphi}{\vec\chi}}
+\frac{2}{3}\lambda_{{\vec\varphi}{\vec\chi}}^2+\frac{1}{6}(\lambda_{\varphi{\vec\varphi}}\lambda_{\varphi{\vec\chi}}
+\lambda_{\chi{\vec\chi}}\lambda_{\chi{\vec\varphi}})\right).
\end{equation}
In the last section, we analyze the theory with $\lambda_{\vec\varphi}=\lambda_{\vec\chi}=\lambda_{{\vec\varphi}{\vec\chi}}\equiv\lambda_t$ and $g_{1t}=g_{2t}\equiv g_t$, and all the other couplings set to zero which describes the pure $f-$wave superconducting transition. We also notice that in this limit, all the three vertex corrections above are equal, as they should be for the consistency, and yield
\begin{equation}\label{vertex-correction-triplet}
\delta\lambda_t=-\frac{7}{3} \lambda_t^2\frac{N_d}{\epsilon}.
\end{equation}

\subsection{Vertex corrections for the singlet-triplet mixed couplings}

Renormalization of the mixed singlet-triplet vertices in Lagrangian $L_b^{st}$ defined in Eq.\ (6) in the main text come from the pairs of couplings in which each member contains a field whose vertex is renormalized. For instance, the $\lambda_{\varphi{\vec\varphi}}$ is renormalized by $\lambda_{\varphi{\vec\varphi}}^2,\lambda_\varphi\lambda_{\varphi{\vec\varphi}},\lambda_{\vec\varphi}\lambda_{\varphi{\vec\varphi}},
\lambda_{\varphi\chi}\lambda_{\chi{\vec\varphi}}$, and $\lambda_{\varphi{\vec\chi}}\lambda_{{\vec\varphi}{\vec\chi}}$ yielding
\begin{equation}
\delta\lambda_{\varphi{\vec\varphi}}=-\frac{N_d}{\epsilon}\left[\lambda_{\varphi{\vec\varphi}}\left(\frac{2}{3}\lambda_{\varphi{\vec\varphi}}
+\frac{5}{6}\lambda_{\vec\varphi}+\frac{1}{2}\lambda_\varphi\right)+\frac{1}{6}\lambda_{\varphi\chi}\lambda_{\chi{\vec\varphi}}+
\frac{1}{2}\lambda_{\varphi{\vec\chi}}\lambda_{{\vec\varphi}{\vec\chi}}\right].
\end{equation}
The correction for the $\lambda{\chi{\vec\chi}}$-vertex is obtained by substituting ${\varphi}\leftrightarrow\chi$ and ${\vec\varphi}\leftrightarrow{\vec\chi}$. The $\lambda_{\chi{\vec\varphi}}$-vertex correction is obtained analogously and reads
\begin{equation}
\delta\lambda_{\chi{\vec\varphi}}=-\frac{N_d}{\epsilon}\left[\lambda_{\chi{\vec\varphi}}\left(\frac{2}{3}\lambda_{\chi{\vec\varphi}}
+\frac{5}{6}\lambda_{\vec\varphi}+\frac{1}{2}\lambda_\chi\right)+\frac{1}{6}\lambda_{\varphi\chi}\lambda_{\varphi{\vec\varphi}}+
\frac{1}{2}\lambda_{\chi{\vec\chi}}\lambda_{{\vec\varphi}{\vec\chi}}\right].
\end{equation}
Correction for the $\lambda_{\varphi{\vec\chi}}$-vertex is obtained from above by replacing ${\varphi}\leftrightarrow\chi$ and ${\vec\varphi}\leftrightarrow{\vec\chi}$.
\subsection{Renormalization conditions and beta-functions for the quartic couplings}

After having obtained all the vertex corrections and the bosonic field-strength renormalizations, one can now write the renormalization conditions for
the quartic couplings which then yield the beta functions. For a coupling $\lambda_{\alpha\beta}$ the renormalization condition reads (no summation over repeated indices is assumed here)
\begin{equation}
Z_\alpha Z_\beta \mu^{-\epsilon} \lambda_{\alpha\beta,0}+{\tilde\delta}\lambda_{\alpha\beta}+{\delta}\lambda_{\alpha\beta}=\lambda_{\alpha\beta}.
\end{equation}
These renormalization conditions, together with the wavefunction renormalization for the bosonic fields and vertex corrections computed in Secs.\ IIIA,B,C yield the following beta functions for the quartic couplings near the critical point describing the transition to $f+is$ state, after setting $g_{1s}=g_{2t}=0$, $\lambda_\varphi=\lambda_{\vec\chi}=\lambda_{\varphi\chi}=\lambda_{{\vec\varphi}{\vec\chi}}=0$, as well as $\lambda_{\varphi{\vec\varphi}}=\lambda_{\chi{\vec\chi}}=\lambda_{\varphi{\vec\chi}}=0$
\begin{eqnarray}
\beta_{\lambda_{\chi}}&=&\epsilon\lambda_\chi-4Ng_{2s}^2(\lambda_\chi-6g_{2s}^2)-\frac{1}{2}(3\lambda_\chi^2+\lambda_{\chi{\vec\varphi}}^2), \nonumber \\
\beta_{\lambda_{{\vec\varphi}}}&=&\epsilon\lambda_{{\vec\varphi}}-4Ng_{2t}^2(\lambda_{{\vec\varphi}}-6g_{2t}^2)-\frac{1}{6}(11\lambda_{{\vec\varphi}}^2
+\lambda_{\chi{\vec\varphi}}^2),\nonumber \\
\beta_{\lambda_{\chi{\vec\varphi}}}&=&\epsilon\lambda_{\chi{\vec\varphi}}-2N\lambda_{\chi{\vec\varphi}}(g_{2s}^2+g_{1t}^2)+24Ng_{2s}^2 g_{1t}^2
-\frac{1}{2}\lambda_{\chi{\vec\varphi}}(\lambda_{\chi}+\lambda_{{\vec\varphi}})-\frac{2}{3}\lambda_{\chi{\vec\varphi}}^2,
\end{eqnarray}
with $(g_{1t},g_{2s})\equiv (g^\ast_{1t},g^\ast_{2s})$ with $(g^\ast_{1t},g^\ast_{2s})$ as the critical Yukawa couplings in Eq.\ (8) in the main text, and rescaling $N_d\lambda\rightarrow \lambda$ applied for each quartic coupling. The fixed points for the quartic couplings cannot be expressed in the closed form for an arbitrary number of four-component Dirac fermions. A numerical analysis in case $N=2$ as in graphene then yields the stable fixed point for $(\lambda^\ast_{\chi},\lambda^\ast_{{\vec \varphi}},\lambda^\ast_{\chi{\vec\varphi}})=(0.972,0.992,1.103)\epsilon$, which is the result quoted in the main text.

\section{Correlation-length exponents and anomalous dimensions at the $f+is$ multicritical point}

Correlation-length exponent corresponding to the two relevant mass parameters, $m_\chi$ and $m_{\vec\varphi}$ close to the $f+is$ critical point is determined using the corrections that these parameters get from the quartic bosonic couplings which are obtained in the standard way\cite{justin} and yield the following renormalization conditions
\begin{eqnarray}
&&Z_\chi m_{\chi,0}^2-m_\chi^2\frac{1}{2}(\lambda_{\chi}+\lambda_{\chi{\vec\phi}})\frac{N_d}{\epsilon}=m_\chi^2,\\
&&Z_{\vec\varphi} m_{{\vec\varphi},0}^2-m_{\vec\varphi}\frac{1}{6}(5\lambda_{\vec\varphi}+\lambda_{\chi{\vec\varphi}})\frac{N_d}{\epsilon}=m_{\vec\varphi}^2.
\end{eqnarray}
These renormalization conditions in turn yield the beta functions for the masses
\begin{eqnarray}
&&\beta_{m_\chi^2}=\left(2-\frac{1}{2}(\lambda_{\chi}+\lambda_{\chi{\vec\varphi}})-2N g_{2s}^2\right)m_{\chi}^2\equiv \nu_\chi^{-1}m_{\chi}^2\\
&&\beta_{m_{\vec\varphi}^2}=\left(2-\frac{1}{6}(5\lambda_{\vec\varphi}+\lambda_{\chi{\vec\varphi}})-2N g_{1t}^2\right)m_{\vec\varphi}^2\equiv \nu_{\vec\varphi}^{-1}m_{\chi}^2.
\end{eqnarray}
These beta functions yield the correlation-length exponents
\begin{eqnarray}
&&\nu_\chi=\frac{1}{2}+\frac{1}{8}(\lambda_{\chi}^*+\lambda_{\chi{\vec\varphi}}^*)+\frac{N}{2}(g_{2s}^*)^2\\
&&\nu_{\vec\varphi}=\frac{1}{2}+\frac{1}{24}(5\lambda_{\vec\varphi}^*+\lambda^*_{\chi{\vec\varphi}})+\frac{N}{2}(g_{1t}^*)^2,
\end{eqnarray}
which, after substituting the couplings at the multicritical point, yield the values of the correlation-length exponent, reported in the main text.

Anomalous dimensions for the Dirac fermions and the OP fields in the vicinity of the $f+is$ critical point are readily obtained from the corresponding field-strength renormalization constants (\ref{Z-bosonic}) and (\ref{Z-Psi})
\begin{equation}
\eta_\Psi=\frac{1}{2}\left((g^*_{1s})^2+3(g_{1t}^*)^2\right),\,\,\eta_\chi=2N(g_{2s}^*)^2,\,\,\eta_{\vec\varphi}=2N(g_{1t}^*)^2,\eta_\varphi=\eta_{\vec\chi}=0.
\end{equation}

\section{Critical theory for the transition into the triplet $f$-wave pairing state}

As already announced, our results enable us to address the critical theory describing the quantum phase transition of Dirac fermions into the triplet $f$-wave superconducting state. In this case the real and the imaginary OPs in the triplet channel are the only critical bosonic degrees of freedom, and thus we set all the couplings in the singlet channel, as well as the mixed singlet-triplet couplings to zero. The remaining Yukawa couplings we set equal $g_{1t}=g_{2t}\equiv g_t$ as well as all the remaining quartic bosonic couplings, $\lambda_{\vec\varphi}=\lambda_{\vec\chi}\equiv\lambda_t$. Using Eqs.\ (\ref{quartic-Yukawa correction}) and (\ref{vertex-correction-triplet}), the corresponding renormalization conditions for these two coupling constants read (notice that in this case Yukawa vertex does not receive any vertex correction, same as near the s-wave criticality\cite{RJH-RC1})
\begin{eqnarray}
Z_{\vec\varphi}^{1/2}Z_\Psi g_{t,0}=g_t,\\
Z_{\vec\varphi}^2\lambda_{t,0}+24N g_t^2\frac{N_d}{\epsilon}-\frac{7}{3}\lambda_t^2\frac{N_d}{\epsilon}=\lambda_t.
\end{eqnarray}
Using these equations and the wavefunction renormalizations (\ref{Z-bosonic}) and (\ref{Z-Psi}), after rescaling the coupling with $N_d$, we obtain the infrared beta functions
\begin{eqnarray}
\beta_{g_t^2}&=&\epsilon g_t^2-2(N+3)g_t^4,\\
\beta_{\lambda_t}&=&\epsilon \lambda_t -\frac{7}{3}\lambda_t^2-4Ng_t^2(\lambda_t-6g_t^2).
\end{eqnarray}
These beta functions have only one fully stable fixed point in the critical plane (all masses are set to zero) which is located at
\begin{equation}
(g_{1t}^*)^2=\frac{\epsilon}{2(N+3)},\;\;\;\;\; \lambda_t^*=\frac{3\epsilon}{14(N+3)}\left(\sqrt{(N-3)^2+56N}-(N-3)\right),
\end{equation}
and since for any number of Dirac fermion flavours, the critical quartic coupling is positive, the transition to the triplet $f$-wave superconducting state is always of the second order. The correlation length exponent can be readily found from the renormalization condition
\begin{equation}
Z_{\vec\varphi}m^2_{{\vec\varphi},0}-\frac{4}{3}\lambda_t m_{\vec\varphi}^2=m_{\vec\varphi}^2,
\end{equation}
yielding
\begin{equation}
\nu_t=\frac{1}{2}+\frac{1}{3}\lambda_t^*+\frac{N}{2}(g_t^*)^2.
\end{equation}
Finally, anomalous dimensions are readily obtained from the fermionic and bosonic wavefunction renormalizations (\ref{Z-bosonic}) and (\ref{Z-Psi}) and at the critical point are
\begin{equation}
\eta_\Psi=\frac{3\epsilon}{2(N+3)},  \;\;\;\;    \eta_{\vec\varphi}=\frac{N\epsilon}{N+3}.
\end{equation}
These critical exponents are different from the ones obtained near the $s$-wave quantum-critical point.\cite{RJH-RC1}


\begin{thebibliography}{}
\bibitem{wekcouplingexplanation} Repulsive interactions could be overtaken by the pairings in strained graphene near the Dirac point in proximity of a superconductor, since they remain weak upon integrating out fast Fourier modes within the momentum shell $(1/a,1/l_b)$, where $l_b$ is the axial magnetic length.
\end{thebibliography}
\end{document}